\documentclass[12pt,a4paper]{article}
\usepackage[T1]{fontenc}        
\usepackage{lmodern}            
\usepackage[ngerman,english]{babel}
\usepackage{amsmath}
\usepackage{amssymb}
\usepackage{amsfonts}
\usepackage{setspace}           
\usepackage[dvips]{graphicx}
\title{
  {\vspace{-2cm} \normalsize
     \hfill MS-TP-10-05
  }\\[5mm]
Simulation of $4d$ $\mathcal{N}=1$ supersymmetric Yang-Mills theory 
with Symanzik improved gauge action and stout smearing}
\author{K.\ Demmouche$^{a}$,
        F.\ Farchioni$^{a}$,
        A.\ Ferling$^{a}$,
        I.\ Montvay$^{b}$, \\
        G.\ M\"unster$^{a}$\thanks{email: munsteg@uni-muenster.de},
        E.E.\ Scholz$^{c}$,
        J.\ Wuilloud$^{a}$\\[5mm]
 {\small $^a$ Universit\"at M\"unster,
  Institut f\"ur Theoretische Physik,}\\
 {\small Wilhelm-Klemm-Str.~9, D-48149 M\"unster,
  Germany}\\
 {\small $^b$ Deutsches Elektronen-Synchrotron DESY, Notkestr.\,85,
  D-22603 Hamburg, Germany}\\
 {\small $^c$ Universit\"at Regensburg, 
  Institut f\"ur Theoretische Physik,}\\
 {\small Universit\"atsstr.~31, D-93040 Regensburg,
  Germany}\\[5mm]}
\date{March 10, 2010}
\newcommand{\U}{\mathrm{U}}
\newcommand{\SU}{\mathrm{SU}}
\newcommand{\Tr}{\mbox{Tr}}

\newcommand{\I}{\ensuremath{\mathrm{i}}}
\newcommand{\E}{\ensuremath{\mathrm{e}}}
\begin{document}
\maketitle

\begin{abstract}
We report on the results of a numerical simulation concerning the low-lying
spectrum of four-dimensional $\mathcal{N}=1$ SU(2) Supersymmetric Yang-Mills
(SYM) theory on the lattice with light dynamical gluinos. In the gauge
sector the tree-level Symanzik improved gauge action is used, while we use
the Wilson formulation in the fermion sector with stout smearing of the
gauge links in the Wilson-Dirac operator. The ensembles of gauge
configurations were produced with the Two-Step Polynomial Hybrid Monte Carlo
(TS-PHMC) updating algorithm. We performed simulations on large lattices up
to a size of $24^3 \cdot 48$ at $\beta=1.6$. Using QCD units with the Sommer
scale being set to $r_0=0.5\,{\rm fm}$, the lattice spacing is about
$a\simeq 0.09\,{\rm fm}$, and the spatial extent of the lattice corresponds
to $2.1$ fm. At the lightest simulated gluino mass the spin-1/2 gluino-glue
bound state appeared to be considerably heavier than its expected
super-partner, the pseudoscalar bound state. Whether supermultiplets are
formed remains to be studied in upcoming simulations.
\end{abstract}

\newpage
\section{Introduction}

In recent years supersymmetric theories have aroused increasing interest in
elementary particle physics. The supersymmetric extension of the Standard
Model with $\mathcal{N}=1$ supercharge is considered to be an interesting
candidate for a quantum field theory with phenomenological relevance in the
near future. Supersymmetry (SUSY) is an essential ingredient also for other
models beyond the Standard Model.

The $\mathcal{N}=1$ Supersymmetric Yang-Mills (SYM) theory is the minimal
supersymmetric extension of the $\SU (N_{c})$ gauge theory describing
self-interactions of gauge fields $A_{\mu}^a$, corresponding to the {\em
gluons} ($g$). The supersymmetric partners of the gluons are described by
spin-1/2 Majorana fermion fields $\lambda^{a}\; (a=1,\dots, N_{c}^2-1)$, the
{\em gluinos} ($\tilde g$). Compatibility of SUSY with gauge invariance
requires that the gluinos transform in the adjoint representation of the
gauge group. This theory describes the interactions between gluons and
gluinos. The Lagrangian of Euclidean SYM theory in the continuum, including
a SUSY breaking mass term, reads
\begin{equation}
\label{eq:SYM-Lag}
\mathcal{L}_{\textrm{SYM}} 
= \frac{1}{4} F_{\mu\nu}^a F_{\mu\nu}^a
+ \frac{1}{2} {\bar\lambda}^a \gamma_{\mu} (\mathcal{D}_{\mu}\lambda)^a
+ \frac{m_{\tilde g}}{2}\, {\bar\lambda}^a\lambda^a \,,
\end{equation}
where $\mathcal{D}_{\mu}$ denotes the gauge covariant derivative in the
adjoint representation. The gluino mass term introduces a soft breaking of
supersymmetry.

In the low-energy regime the interactions become strong. Arguments based on
the low-energy effective Lagrangian approach \cite{VeYa,FaGaSch} predict the
occurrence of non-perturbative dynamics like confinement and spontaneous
chiral symmetry breaking in SUSY gauge theories. Confinement is realised by
colourless bound states. Since both gluons and gluinos transform according
to the adjoint representation, bound states can be built by any number of at
least two gluons and gluinos. In the case where the last term in
Eq.~(\ref{eq:SYM-Lag}) is switched off ($m_{\tilde g}=0$), an anomalous
global chiral symmetry $\U (1)_{\lambda}$ is present. This symmetry is
equivalent to the $R$-symmetry in supersymmetric models. The anomaly does
not break the global chiral symmetry completely and a discrete subgroup
$Z_{2N_{c}}$ remains. As in the case of QCD, the discrete chiral symmetry is
expected to be spontaneously broken to $Z_{2}$ by the non-vanishing value of
the gluino condensate $\langle {\bar\lambda} \lambda\rangle$. The
consequence of this spontaneous breaking is the existence of $N_{c}$
degenerate ground states with different orientations of the gluino
condensate.

SYM is also equivalent to QCD with a single quark flavour ($N_{f}=1$ QCD) in
the limit of a large number of colours ($N_{c}\rightarrow \infty$), where
the Majorana spinor is replaced by a single Dirac spinor in the
antisymmetric representation of the gauge group
\cite{ArmoniShifmanVeneziano}. The latter model is also object of
investigation by our collaboration \cite{nf1}.

Since confinement occurs in low-energy SYM, standard analytical methods like
perturbation theory fail and non-perturbative methods are required. This
motivates the introduction of the lattice formulation of SYM. The first
lattice formulation of SYM suitable for numerical simulations has been
proposed by Curci and Veneziano \cite{CuVe}. It it based on the Wilson
discretisation, which proved to be successful in lattice QCD computations in
spite of its known limitations. First non-perturbative investigations of SYM
on the lattice using this formulation have been performed by \cite{DoetAl}
in the quenched approximation, and by the DESY-M\"unster-Roma collaboration
with dynamical fermions; see Ref.~\cite{Montvay_SYM} for a review, and
references
\cite{CaetAl,FaetAl,fed-peetz-res,Demmouche:2008aq,Demmouche:2008ms}. SUSY
is broken explicitly by the lattice discretisation. Additionally, in the
Wilson approach the mass term and the Wilson-term break both chirality and
SUSY explicitly. Both symmetries are expected to be recovered in the
continuum limit by tuning the relevant bare mass term to its critical value
corresponding to a massless gluino ($m_{\tilde g}=0$), and the gauge
coupling towards zero.

In recent years, simulations of $\mathcal{N}=1$ SYM on the lattice using
Ginsparg-Wilson fermions with good chiral properties, such as domain wall
fermions, have been initiated
\cite{Fleming:2000fa,Endres:2008tz,Endres:2009yp,Giedt:2008xm}. For large
lattice volumes and small lattice spacings these formulations require,
however, a significantly larger amount of computing resources than the
Wilson formulation.  The gain of no need for tuning the position of the zero
gluino mass point does not compensate by far the advantage of Wilson
fermions.

In the past, investigations of the gluino dynamics have been performed using
the Two-Step Multi Bosonic (TSMB) algorithm \cite{MontvayTSMB}. This
algorithm was developed in the framework of the DESY-M\"unster
collaboration. Recently, the Two-Step Polynomial Hybrid Monte Carlo
(TS-PHMC) algorithm \cite{MontvayPHMC} has been developed and implemented
for SYM. This algorithm offers more efficiency and improvements compared to
the TSMB algorithm and allows us to collect higher statistics and to
simulate small gluino masses $m_{\tilde g}$ in this study. Furthermore, due
to available computer resources we simulated the theory on volumes with
extension larger than 2 fm, which is expected to be the minimally required
volume for spectroscopic studies.

The main purpose of this work is to continue the project of the
DESY-M\"unster collaboration for the simulation of $\mathcal{N}=1$ $\SU (2)$
SYM. We present new accurate results obtained with the newly used TS-PHMC
algorithm and improved actions.

The most important characteristics of the theory is the mass spectrum of
bound states, for which the low-energy effective theories predict a
reorganisation of the masses in two massive Wess-Zumino supermultiplets at
the SUSY point \cite{VeYa,FaGaSch}, where the soft breaking vanishes. The
introduction of a small gluino mass removes the mass degeneracy between the
supermultiplet members. In the lower supermultiplet the ordering of the
states with increasing mass is: scalar glueball $0^{++}$, spin-1/2
gluino-glueball ($\chi_{L}$), pseudoscalar glueball $0^{-+}$. The ordering
is reversed in the higher supermultiplet which contains: adjoint
pseudoscalar meson $a$-$\eta^\prime$, spin-1/2 gluino-glueball ($\chi_{H}$),
adjoint scalar meson $a$-$f_{0}$.

The plan of this paper is as follows: in the next Section we review the
lattice formulation and describe the simulation details.
Section~\ref{sec:confinement} is devoted to the static quark potential and
the determination of the scale. Methods for the determination of the masses
of bound states are described in Section~\ref{sec:masses}. In
Section~\ref{sec:spectrum} the results on the spectrum are collected and
discussed. Finally, we conclude our findings in
Section~\ref{sec:conclusion}.

\section{Lattice formulation of $\mathcal{N}=1$ SYM theory}
\label{sec:lattice}

The Curci-Veneziano action of $\mathcal{N}=1$ $\SU (2)$ SYM theory on a
lattice, $S = S_{g} + S_{\tilde g}$, contains the usual plaquette gauge
field action $S_{g}$, and a fermionic action $S_{\tilde g}$ for the gluino.
The gauge action $S_{g}$ can be extended to a more general form which
includes, besides the usual $(1 \times 1)$ Wilson loop plaquette term,
$(1\times 2)$ Wilson loops of perimeter six. We employ the {\em tree-level
improved Symanzik} (tlSym) gauge action, given for $\SU (N_c)$ colour group
by
\begin{equation}
S_{g}^{\textrm{tlSym}} = \beta\sum_{x} \left(c_{0} \sum_{\mu<\nu;\,\mu,\nu=1}^4
\left\{ 1 - \frac{1}{N_{c}}\,{\rm Re\,} U_{x\mu\nu}^{1\times 1} \right\}
+ c_{1} \sum_{\mu\ne\nu;\,\mu,\nu=1}^4
\left\{ 1 - \frac{1}{N_{c}}\,{\rm Re\,} U_{x\mu\nu}^{1\times 2}\right\}
\right),
\end{equation}
with the normalisation condition $c_{0}=1-8c_{1}$. The bare gauge coupling
$g_{0}$ is related to the lattice parameter $\beta$ by the usual relation
$\beta = 2N_{c}/g_{0}^2$. For the tlSym action we have $c_{1}=-1/12$
\cite{WeiszWohlert}.

The gluinos are represented by Majorana fermions $\lambda^a$ in the adjoint
representation. They satisfy the Majorana condition
\begin{equation}
\label{eq:majorana-condition}
\lambda = \lambda^\mathcal{C} = \mathcal{C}{\bar\lambda}^T\,,
\end{equation}
where $\mathcal{C} = \gamma_{0} \gamma_{2}$ is the charge conjugation matrix
in the spinorial representation.

In the gluino sector, the Wilson formulation for fermions proposed in
\cite{CuVe} introduces the Wilson term proportional to $r$, which is an
irrelevant term in the continuum limit. We set the Wilson parameter to
$r=1$. The fermion part $S_{\tilde g}$ of the action is then given by
\begin{equation}
S_{\tilde g}
= \frac{1}{2} \sum_x \bar{\lambda}(x)\lambda(x) 
- \frac{\kappa}{2} \sum_x \sum_\mu
[\bar{\lambda}(x + \hat{\mu}) V_\mu(x) (1 + \gamma_\mu) \lambda(x)
+ \bar{\lambda}(x) V_\mu^T(x) (1 - \gamma_\mu) \lambda(x+\hat{\mu})] \,,
\end{equation}
where $\kappa$ is the bare hopping parameter which encodes the bare gluino
mass $\kappa=(2m_{{\tilde g},0}+8)^{-1}$.  The real orthogonal matrices
$V_{\mu}(x)$ are the gauge links in the adjoint representation:
\begin{equation} 
[V_\mu(x)]^{ab} 
\equiv 2 \Tr [U_{\mu}^{\dag}(x) T^a U_{\mu}(x) T^b] 
= [V_\mu^*(x)]^{ab} = [V_{\mu}^{-1}(x)]^{ba}\,, 
\end{equation}
where $T^a$ are the generators of SU($N_c$) satisfying
$2\Tr(T^{a}T^{b})=\delta^{ab}$. In case of SU(2) one has
$T^a=\frac{1}{2}\sigma^a$ with the Pauli matrices $\sigma^a$.

The links $U_{x,\mu}$ in the fermion action can be replaced by {\em
stout}-smeared links \cite{MorningstarPeardon}. This has the advantage that
short range topological defects of the gauge field and the corresponding
small eigenvalues of the fermion matrix are removed. Both the tlSym gauge
action and the stout smeared links in the fermionic part of the lattice
action are introduced in order to accelerate the approach to the continuum
limit as $\beta\to\infty$.

The stout smeared links are defined by
\begin{equation}
U^{(1)}_{x,\mu} \equiv U_{x,\mu}\,\exp\left\{
\frac{1}{2}\left( \Omega_{x,\mu} - \Omega^\dagger_{x,\mu} \right) -
\frac{1}{2N_c}{\rm\,Tr}
\left( \Omega_{x,\mu} - \Omega^\dagger_{x,\mu} \right)
\right\}.
\end{equation}
Here $U_{x,\mu}$ denotes the original ``thin'' gauge links, and
\begin{equation}
\Omega_{x,\mu} \equiv \rho\, U^\dagger_{x,\mu}\, C_{x,\mu}
\end{equation}
with the sum of ``staples''
\begin{equation}
C_{x,\mu} \equiv \sum_{\nu\ne\mu} \left(
U^\dagger_{x+\hat{\mu},\nu} U_{x+\hat{\nu},\mu} U_{x,\nu} +
U_{x-\hat{\nu}+\hat{\mu},\nu} U_{x-\hat{\nu},\mu} U^\dagger_{x-\hat{\nu},\nu}
\right).
\end{equation}
$\rho$ is an arbitrary parameter which we fix in this work to $\rho=0.15$.
In principle, the smearing defined by the above equations can be iterated
several times, but then the fermion action becomes extended over a larger
region on the lattice. We prefer to keep the action well localised and hence
only perform a single smearing step.

Writing the gluino action as
\begin{equation}
S_{\tilde g}
= \frac{1}{2} \sum_{xy} a^4\, {\bar\lambda}_y Q_{yx} \lambda_x\,,
\end{equation}
$Q$ is the non-hermitian fermion matrix or lattice Wilson-Dirac operator for
Dirac fermions in the adjoint representation. Using relation
(\ref{eq:majorana-condition}), the fermion action can be rewritten in terms
of the antisymmetric matrix $M= \mathcal{C} Q$. Integration of the fermionic
variables yields the Pfaffian of $M$, 
\begin{equation}
\int\!\!\mathcal{D}\lambda\, \E^{-S_{\tilde g}}
= \textrm{Pf}\,(M),
\end{equation}
whose absolute value equals the square
root of the fermion determinant:
\begin{equation}
|\textrm{Pf}\,(M)| = \sqrt{\det(M)} = \sqrt{\det(Q)}\,.
\end{equation}
Effectively, this corresponds to a flavour number $N_{f}=1/2$. In the Wilson
setup, $\det(Q)$ and $\det(M)$ are always real and positive, but the
Pfaffian $\textrm{Pf}\,(M)$ can become negative even for positive gluino
masses.

In our numerical simulations we include the dynamics of the gluino by the
Two-Step Polynomial Hybrid Monte Carlo (TS-PHMC) \cite{MontvayPHMC}
algorithm with flavour number $N_{f}=1/2$. This has the consequence that
only the absolute value of the Pfaffian is taken into account in the
updating of the gauge field configuration. The sign of the Pfaffian has to
be included in a reweighting step when calculating expectation values. It
can be shown that the sign of the Pfaffian is equal to the sign of the
product of half of the doubly degenerate negative real eigenvalues of $Q$.
For positive gluino masses sufficiently far away from zero, a negative sign
of the Pfaffian rarely occurs in the updating sequence and therefore in this
situation a sign problem does not show up. Approaching the limit of
vanishing gluino mass we monitor the sign of the Pfaffian and take it into
account by reweighting. It turned out that only in our runs $D$ and $D_s$
(see Table \ref{tab:setup} below) a noticeable number of configurations with
negative sign occured; the highest fraction being in point $D_s$, where they
amount to 3\,\% of all configurations. The effect of the negative signs on
the particle masses turned out to be negligible.

The parameters of the $\mathcal{N}=1$ SYM on the lattice are the lattice
gauge coupling $\beta$ and the fermionic hopping parameter $\kappa$.
Similarly to QCD, the mass term proportional to $m_{{\tilde g},0}$ breaks
chirality explicitly. In the present case it also breaks the supersymmetry.
A massless gluino, $m_{\tilde g}=0$, is obtained by tuning the bare mass
term to its critical value ($m_{{\tilde g},0} \rightarrow m_{c}$) or
equivalently $\kappa\rightarrow\kappa_{c}$.

\begin{table}[ht]
\caption{\em Algorithmic parameters for TS-PHMC runs with tlSym gauge action
at $\beta=1.6$. Runs labelled with subscript $s$ have been performed with
Stout-links. $N_{\textrm{conf}}$ is the number of configurations produced,
$r_{0}$ is the Sommer scale parameter, $am_{\pi}$ is the adjoint pion mass
in lattice units, and $M_{r}$ is the dimensionless quantity $M_{r}\equiv
(r_{0}m_{\pi})^2$ used to estimate the gluino mass. In $M_{r}$ the values of
$r_{0}/a$ extrapolated to $\kappa_{c}$ have been used.}
\begin{center}
\begin{tabular}{ll|llrlll}
Run & $L^3.T$ & $\beta$ & $\kappa$ & $N_{\textrm{conf}}$ & $r_{0}/a$ 
   & $am_{\pi}$ & $M_{r}$\\
\hline \hline
$A$  & $16^3\cdot 32$ & 1.6 & 0.1800 & 2500 & 2.9(1) & 1.3087(12) & 45.6(4.2)\\
$B$  & $16^3\cdot 32$ & 1.6 & 0.1900 & 2700 & 3.3(1) & 1.0071(12) & 27.0(2.5)\\
$C$  & $16^3\cdot 32$ & 1.6 & 0.2000 &10847 & 4.242(87) & 0.5008(13) & 6.68(62)\\
$D$  & $16^3\cdot 32$ & 1.6 & 0.2020 & 6947 & 5.04(26) & 0.221(12) & 1.30(19)\\
\hline
$\rule{0pt}{14pt}\bar A$ & $24^3\cdot 48$ & 1.6 & 0.1980 & 1480 & 3.885(63) & 0.6415(13) & 11.0(1.0)\\
$\bar B$ & $24^3\cdot 48$ & 1.6 & 0.1990 & 1400 & 4.16(12) & 0.5759(17) & 8.83(82)\\
$\bar C$ & $24^3\cdot 48$ & 1.6 & 0.2000 & 6465 & 4.33(19) & 0.4947(13) & 6.52(61)\\
\hline
$A_s$ & $24^3\cdot 48$ & 1.6 & 0.1500 & 370 & &0.9469(38) & 28.69(89)\\
$B_s$ & $24^3\cdot 48$ & 1.6 & 0.1550 & 1730 & 4.324(39) & 0.5788(16) & 10.72(33)\\
$C_s$ & $24^3\cdot 48$ & 1.6 & 0.1570 & 2110 & 5.165(88) & 0.3264(23) & 3.41(11)\\
$D_s$ & $24^3\cdot 48$ & 1.6 & 0.1575 & 2260 & 5.561(99) & 0.2015(93) & 1.30(13)\\
\hline
\end{tabular}
\end{center}
\label{tab:setup}
\end{table}

In order to study questions related to supersymmetry, one has to approach
the critical value of the hopping parameter $\kappa=\kappa_{c}$
corresponding to zero gluino mass. This tuning problem can be solved rather
easily by means of the adjoint pion mass $m_{\pi}$. This is the pion mass in
the corresponding theory with two Majorana fermions in the adjoint
representation. It is obtained from the exponential decay of the connected
part of the pseudoscalar meson propagator, see below. The pion is not a
physical particle in the spectrum of the SYM theory, but it can be
unambiguously defined in a partially quenched framework.  To determine the
pion mass is rather easy, in fact it is the easiest mass to determine. As
will be detailed in Sec.~\ref{subsec:massless}, the behaviour of the pion
mass-squared is very closely linear as a function of $1/\kappa$ in the
entire range of gluino masses of interest. On the basis of arguments
involving the OZI-approximation of SYM \cite{VeYa}, the adjoint pion mass is
expected to vanish for a massless gluino. Therefore it is enough to perform
two simulations on relatively small lattices at relatively large gluino
masses, from which $\kappa_{c}$ can be obtained by a linear extrapolation.
Proceeding to larger lattices and smaller gluino masses, this estimate can
be continuously improved without any further simulations. In this way the
interesting range of hopping parameters $\kappa < \kappa_{c}$ for the
investigation of the particle spectrum can be determined.

The values of the gauge coupling parameter $\beta$ can be fixed by
investigating the static potential of an external charge in the fundamental
representation and extracting the Sommer scale parameter $r_0/a$
\cite{Sommer}, as discussed in Sec.~\ref{sec:confinement}. In analogy with
QCD, we set the value of $r_0$ by definition to $r_0=0.5\,{\rm fm}$. In this
way we can use familiar QCD units for physical dimensionful quantities.

As a measure for the gluino mass we define the dimensionless quantity
$M_{r}\equiv (r_{0} m_{\pi})^2$, which is expected to be proportional to the
gluino mass.

A summary of the simulation parameters is given in Table~\ref{tab:setup}.
The simulations are performed on $16^3\cdot32$ and $24^3\cdot48$ lattices.
Extrapolated to $\kappa_{c}$ the lattice spacing amounts to $a\simeq
0.097\,{\rm fm}$ for the unstout ensembles and $a\simeq 0.088\,{\rm fm}$ for
the stout ones, see below. The lattice extension $L\simeq 2.1 -2.3\,{\rm
fm}$ is expected to be large enough to allow control over finite volume
effects on the masses of the bound states.

An issue in lattice simulation is the lightness of the dynamical fermions
which leads to slowing down of the update algorithms.  The TS-PHMC algorithm
turned out to be very efficient in producing short autocorrelations among
the gauge configurations. For instance, in the stout-smeared runs on a
$24^3\cdot 48$ lattice the integrated autocorrelation of the average
plaquette (which belongs to the worst quantities from the point of view of
autocorrelations) did always satisfy $\tau_{int}^{plaq} < 10$. The lightest
adjoint pion mass in our simulations was about 440 MeV. Simulations for
smaller gluino masses and/or finer lattice spacings are going on presently.

\section{Static potential and physical scale}
\label{sec:confinement}

Analogy with QCD suggests that the colour charge is confined in SYM, so that
the particle states are colour-singlets. Moreover, SYM is expected to
confine static quarks as in pure Yang-Mills theory: the static
quark-antiquark potential can not be screened by the dynamical gluinos
transforming in the adjoint representation, and a non-vanishing string
tension arises at large distances.

\begin{figure}[!htb]
\vspace{.8cm}
\centering
\includegraphics[width=0.7\textwidth]{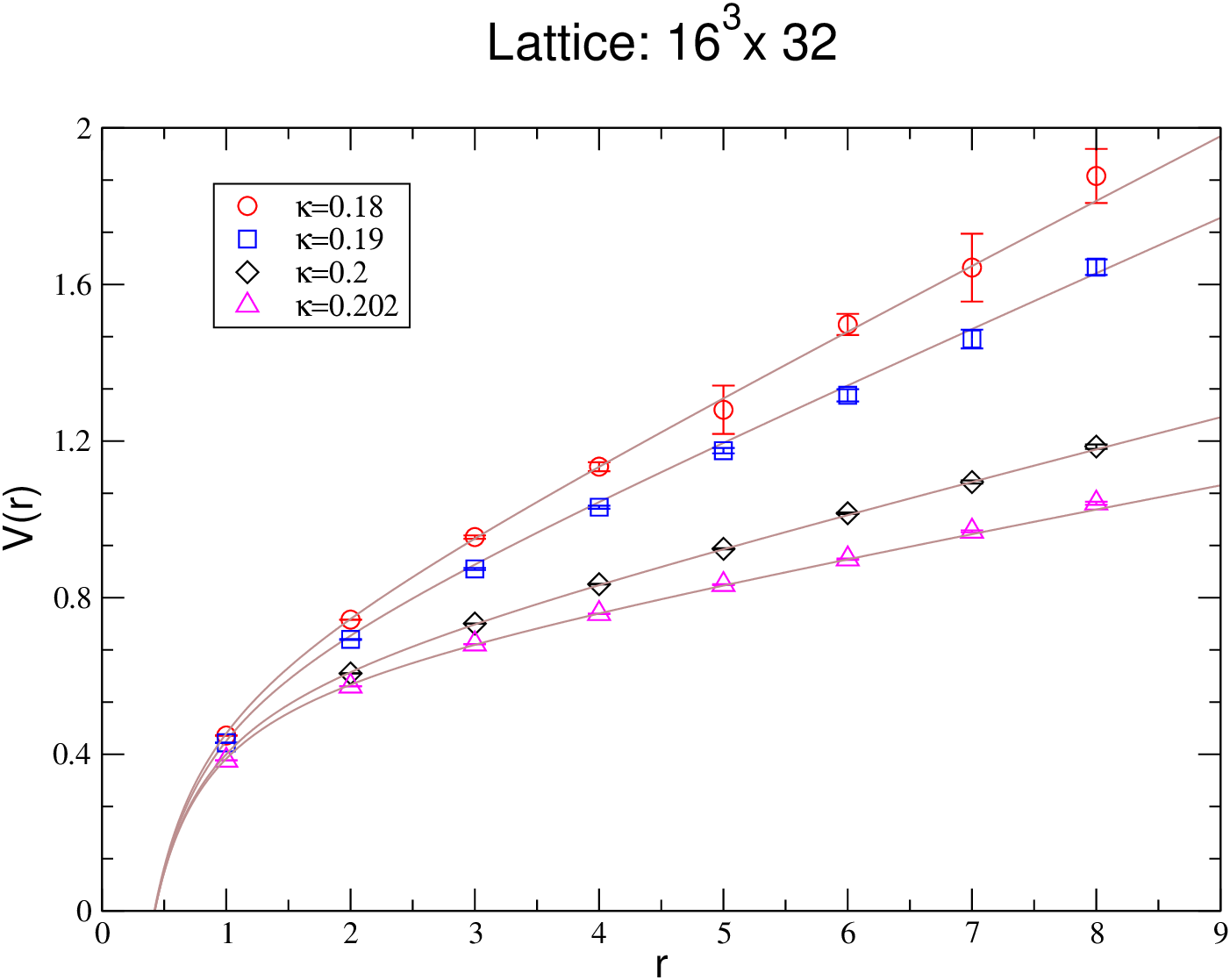}
\parbox[t]{0.8\textwidth}{%
\caption{\label{fig:pot}
The static quark potential in ${\cal N}=1$ SU(2) SYM.
The solid lines are fits to the data.}
}
\end{figure}

The numerical results for the static potential $V(r)$ for runs $A$--$D$ are
shown in Fig.~\ref{fig:pot}. The linear behaviour at large quark-antiquark
separations is compatible with a non-vanishing string tension.

From the behaviour of the static potential at intermediate distances it is
possible to determine the lattice scale, a well-known procedure in lattice
QCD. The scale can be characterised by the Sommer parameter
$r_0$~\cite{Sommer} defined by the relation \begin{equation}
r_0^2\left.\frac{\mbox{d}V}{\mbox{d}r}\right|_{r_0} = 1.65\,. \end{equation}
In the SYM model the string tension could in principle also be used to fix
the scale, but the Sommer scale parameter is more convenient from the
numerical point of view. The numerical results for $r_0/a$ are reported in
the sixth column of Table~\ref{tab:setup}. Following the analogous procedure
in QCD in a mass independent renormalisation scheme, we extrapolate these
data to vanishing adjoint pion mass. For the runs with thin links, where we
combine data from the two volumes, we obtain $r_0/a=5.16(24)$, and for runs
with stout links $r_0/a=5.657(85)$. Using $r_0=0.5\,{\rm fm}$ this
corresponds to $a=0.097$\,fm (thin links) and $a=0.088$\,fm (stout links),
respectively.  The physical size of the simulated boxes is therefore in
these units $L\simeq 1.5 - 2.3\,\mbox{fm}$.

\section{Spectrum of low-lying bound states}
\label{sec:masses}

For the investigation of the spectrum of low-lying bound states we
concentrate on the operators employed for the construction of the low-energy
Lagrangians of~\cite{VeYa} and \cite{FaGaSch}. These are expected to
dominate the dynamics of SYM at low energies. Previous experience on the
determination of low-lying masses is reported in~\cite{CaetAl} and
\cite{fed-peetz-res}. We investigate spin-0 gluino-gluino bilinear operators
(adjoint mesons), a spin-1/2 mixed gluino-glue operator and spin-0 glueball
operators. In some cases smearing techniques such as APE~\cite{Alb87} and
Jacobi smearing~\cite{Jacobi} are applied in order to increase the overlap
of the lattice operator with the low-lying bound state.

\subsection{Adjoint mesons}

The adjoint mesons are colourless states with spin-parity $0^+$ and $0^-$,
composed of two gluinos.  In analogy to flavour singlet states in QCD we
denote the former $a$-$\eta^\prime$ and the latter $a$-$f_0$, where the
prefix $a$ indicates ``adjoint''. The associated projecting operators are
the gluino bilinear operators ${\cal
O}_{\textrm{meson}}=\bar\lambda\Gamma\lambda$ where $\Gamma = 1$ or $\Gamma
= \gamma_5$, respectively. The resulting propagator consists of connected
and disconnected contributions:
\begin{align}
C(x_0-y_0) &= C_{\textrm{conn}}(x_0-y_0) - C_{\textrm{disc}}(x_0-y_0)\nonumber\\
&= \frac{1}{V_s} \sum_{\vec{x}}
\langle \Tr[\Gamma Q^{-1}_{x,y} \Gamma  Q^{-1}_{y,x}]\rangle
- \frac{1}{2V_s} \sum_{\vec{x}} 
\langle \Tr[\Gamma  Q^{-1}_{x,x}] \Tr[\Gamma Q^{-1}_{y,y}]
\rangle,
\end{align}
where $\langle\cdots \rangle$ denotes the average over the gauge sample and
$V_s = L^3$.

The exponential decay of the connected part defines the adjoint pion mass
$m_{\pi}$. This quantity, even if not associated to a physical state of SYM,
can be used to determine the gluino mass, as mentioned in
Sec.~\ref{sec:lattice}. Indeed, according to arguments involving the
OZI-approximation of SYM \cite{VeYa}, the adjoint pion mass is expected to
vanish for a massless gluino and the behaviour $m^2_{\pi} \propto m_{\tilde
g}$ can be assumed for light gluinos~\cite{VeYa,fed-peetz-res}.

As is well known in simulations of QCD, the numerical evaluation of the
disconnected propagators is rather demanding. We employ here two alternative
methods, the Stochastic Estimators Technique (SET) in the spin dilution
variant~\cite{SET}, and the Improved Volume Source Technique (IVST)
\cite{IVST}. As in QCD, the disconnected diagrams are intrinsically noisier
than the connected ones and dominate the level of noise in the total
correlator.

\begin{figure}[!htb]
\vspace{.8cm}
\centering
\includegraphics[width=0.7\textwidth]{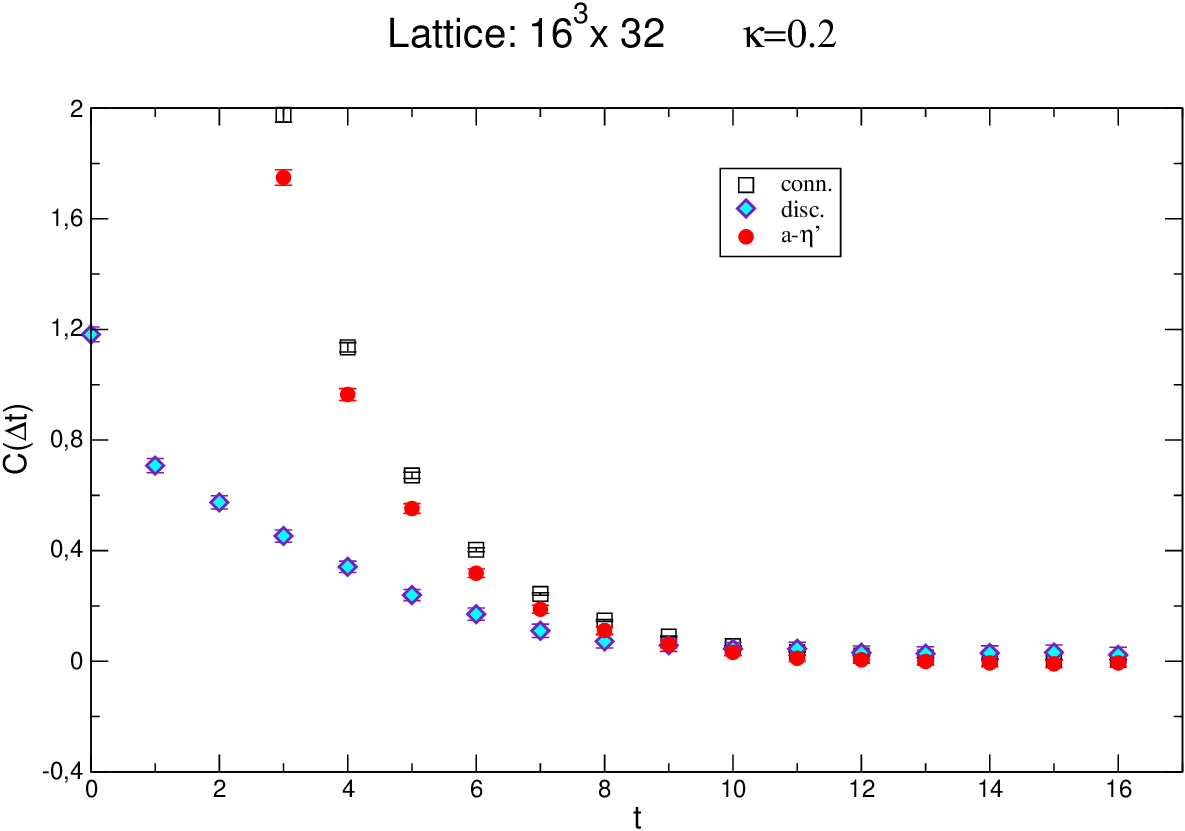}
\parbox[t]{0.8\textwidth}{%
\caption{\label{fig:corr_eta_k02}
Connected and disconnected pieces and
the total time-slice correlation function of the adjoint pseudoscalar
$a$-$\eta^\prime$.}
}
\end{figure}

\begin{figure}[!htb]
\vspace{.8cm}
\centering
\includegraphics[width=0.7\textwidth]{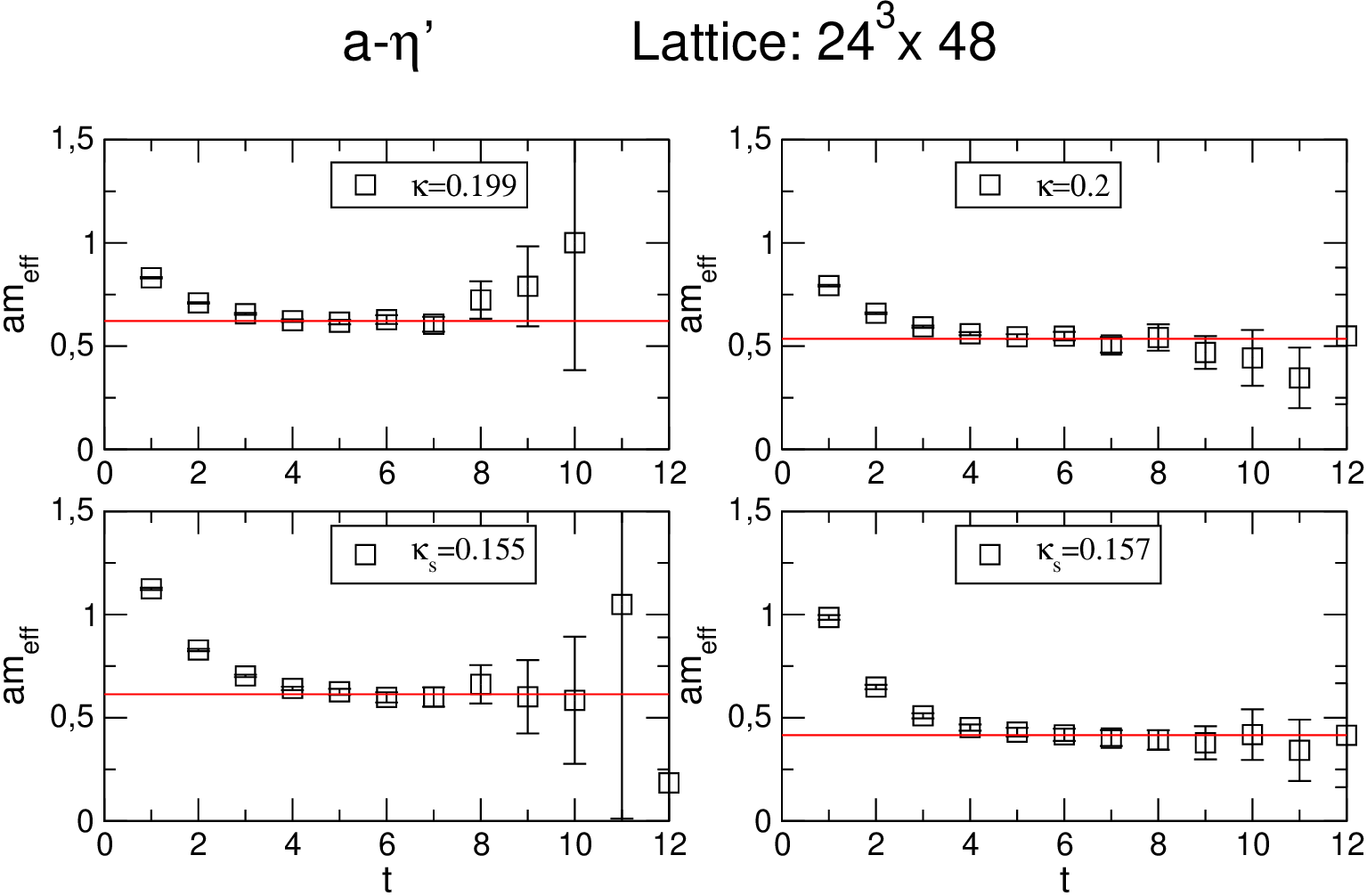}
\parbox[t]{0.8\textwidth}{%
\caption{\label{fig:meff-eta}
Effective mass of the pseudoscalar $a$-$\eta^\prime$.
The horizontal line represents the result from a one-mass-fit.}
}
\end{figure}

In the pseudoscalar channel a reasonable signal-to-noise ratio is obtained,
allowing the extraction of the mass from the mass fit on most samples. As an
example, in Fig.~\ref{fig:corr_eta_k02} we show the result for the
$a$-$\eta^\prime$ correlator for run $C$ together with the two different
contributions. Examples for the effective masses as a function of the time
separation $t$ are shown in Fig.~\ref{fig:meff-eta}

\begin{figure}[!htb]
\vspace{1.1cm}
\centering
\includegraphics[width=0.5\textwidth]{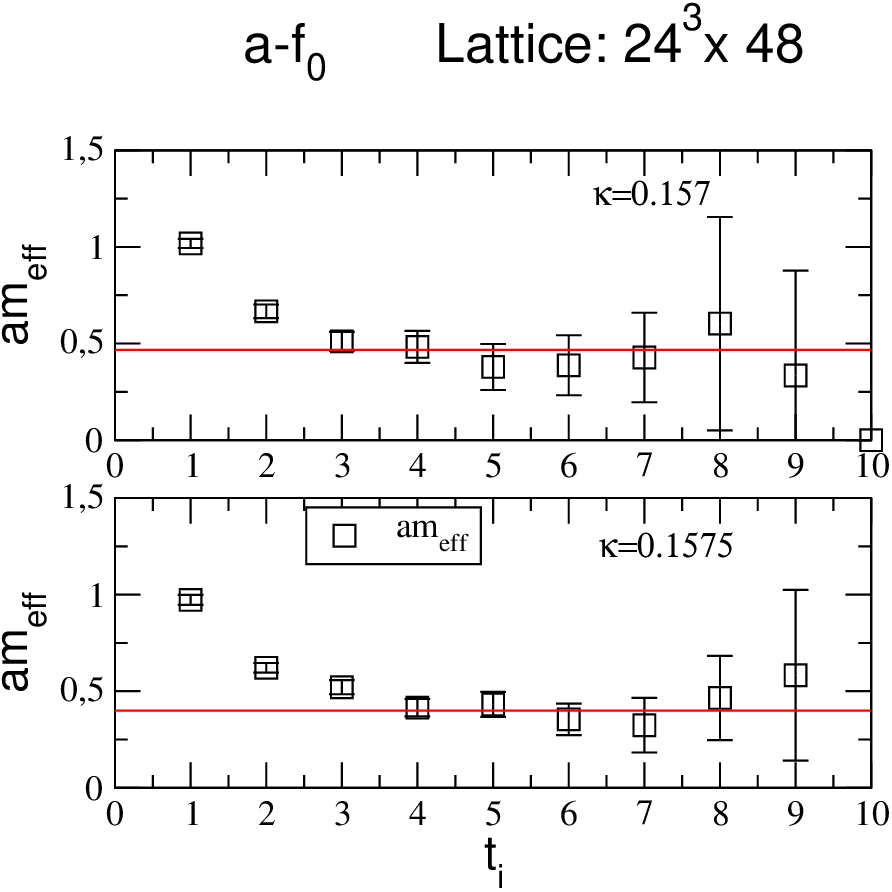}
\parbox[b]{0.8\textwidth}{%
\caption{\label{fig:meff-f0}
Effective mass of the scalar $a$-$f_{0}$. 
The horizontal line represents the result from a one-mass-fit.}
}
\end{figure}

In the scalar channel the extraction of the mass is complicated by the
presence of a vacuum expectation value for the projecting operator $\sim
\langle \bar\lambda\lambda\rangle$. This allowed a relatively precise
determination of the $a$-$f_0$ mass only for the samples with stout
smearing, which give a better signal. Two examples for the effective mass in
this channel are shown in Fig.~\ref{fig:meff-f0}. For the future we plan the
application of variance reduction techniques for a more precise computation
of the disconnected diagrams.

\subsection{Scalar glueball}

As for the adjoint mesons, we investigated the scalar glueball masses also
in both parity channels. In order to improve the signal we applied in this
case APE smearing with the variational method~\cite{Variational}. For the
positive parity glueball $0^+$ we adopted the simplest interpolating
operator built from space-like plaquettes: \begin{equation} {\cal
O}_{\textrm{glue},+}(x) = \Tr_c[U_{12}(x)+U_{23}(x)+U_{31}(x)]\,.
\end{equation} For the negative parity state we considered the eight-link
operator proposed in~\cite{CaetAl}. However, the signal obtained in this
case was too poor to obtain an estimate of the mass. Therefore we restricted
the analysis to the positive parity channel in the following.

\begin{figure}[!htb]
\vspace{1.2cm}
\centering
\includegraphics[width=0.5\textwidth]{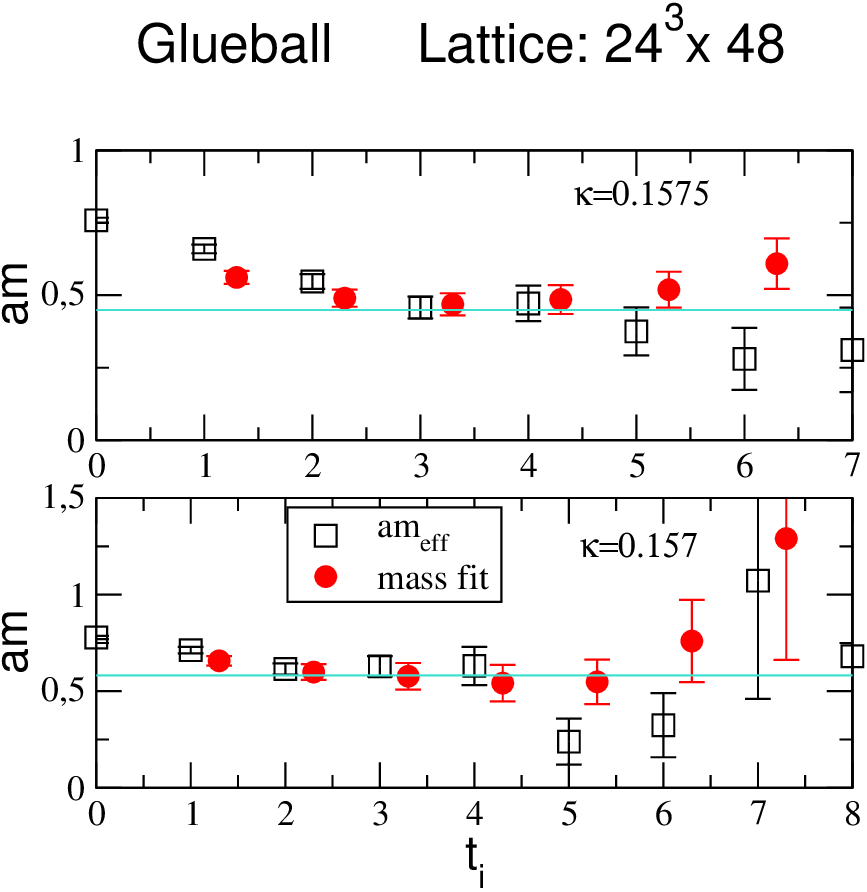}
\parbox[t]{0.8\textwidth}{%
\caption{\label{fig:meff-glueball}
Effective mass and results from one-mass fits
for the scalar glueball $0^+$.}
}
\end{figure}

Also here, as for the scalar $a$-$f_{0}$, the gauge samples generated with
stout links generally turn out to give better results for the glueball
masses. In Fig.~\ref{fig:meff-glueball} two examples of the effective masses
are reported together with the results from one-mass fits with minimal
time-distance $t_i$.

\subsection{Gluino-glueballs}

The gluino-glueballs ($\tilde{g} g$) are spin 1/2 colour singlet states of a
gluon and a gluino. They are supposed to complete the Wess-Zumino
supermultiplet of the adjoint mesons \cite{VeYa}.  For this state we adopt
the lattice version of the gluino-glue operator
$\Tr_c[F\sigma\lambda]$~\cite{VeYa}, where the field-strength tensor
$F_{\mu\nu}(x)$ is replaced by the clover-plaquette operator
$P_{\mu\nu}(x)$~\cite{FaetAl,fed-peetz-res}:
\begin{equation}
{\cal O}_{\tilde{g} g}^{\alpha}(x) = \sum_{i<j}
\sigma_{ij}^{\alpha\beta} \Tr_c[P_{ij}(x)\lambda^{\beta}(x)].
\end{equation}
Here only spatial indices are taken into account in order to avoid links in
the time-direction.  The clover-plaquette operator, having the correct
behaviour under discrete parity and time reversal transformations, is
defined as
\begin{equation} 
P_{\mu\nu}(x) =
\frac{1}{8 \I g_0} \sum_{i=1}^4
\left(U^{(i)}_{\mu\nu}(x) - U^{(i)\dagger}_{\mu\nu}(x)\right) 
\end{equation}
with 
\begin{align} 
U^{(1)}_{\mu\nu} (x) &= U^\dagger_{\nu}(x) U^\dagger_{\mu}(x+ \hat{\nu})
 U_{\nu}(x+\hat{\mu}) U_{\mu}(x)\\
U^{(2)}_{\mu\nu} (x) &= U_{\mu}^\dagger(x) U_{\nu}(x-\hat{\nu}+\hat{\mu})
 U_{\mu}(x-\hat{\nu}) U^\dagger_{\nu}(x-\hat{\nu})\\
U^{(3)}_{\mu\nu} (x) &= U_{\nu}(x-\hat{\nu}) U_{\nu}(x-\hat{\nu}-\hat{\mu})
 U^\dagger_{\mu}(x-\hat{\nu}-\hat{\mu}) U^\dagger_{\mu}(x-\hat{\mu})\\
U^{(4)}_{\mu\nu} (x) &= U_{\mu}(x-\hat{\mu}) U^\dagger_{\nu}(x-\hat{\mu})
 U^\dagger_{\mu}(x+\hat{\nu}-\hat{\mu}) U_{\nu}(x)\,.
\end{align} 

\begin{figure}[!htb]
\vspace{.8cm}
\centering
\includegraphics[width=0.7\textwidth]{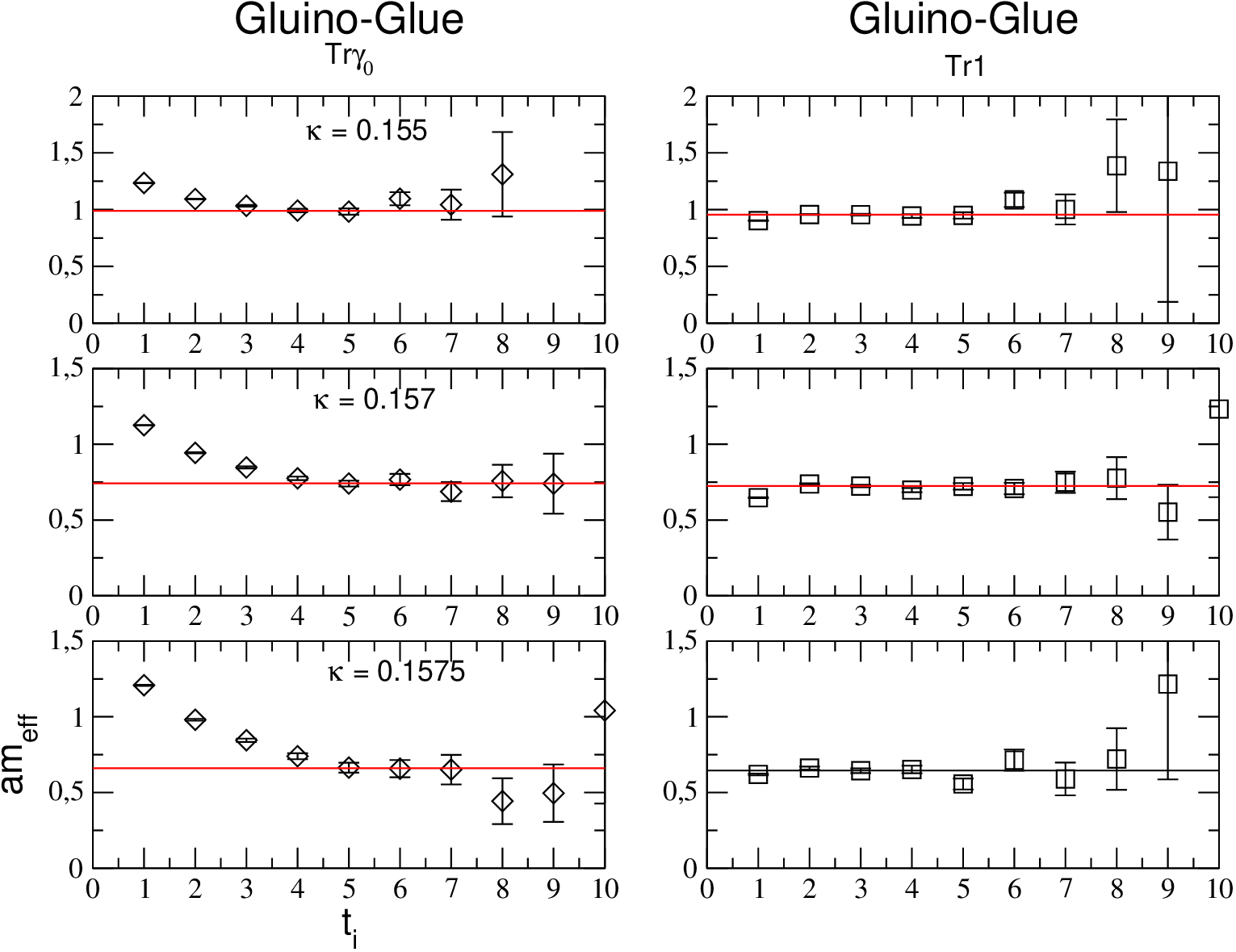}
\parbox[t]{0.8\textwidth}{%
\caption{\label{fig:meff-gluino-glue}
Effective mass of the gluino-glueball $\tilde{g} g$.}
}
\end{figure}

The full correlator of the gluino-glue operator,
\begin{equation}
C^{\alpha\beta}_{\tilde{g} g}(x_0-y_0)
= -\frac{1}{4} \sum_{\vec x, \vec y} \sum_{i,j,k,l}
\left\langle \sigma_{ij}^{\alpha\alpha^\prime} \mbox{Tr}[U_{ij}(x)
\sigma^a] Q^{-1}_{xa\alpha^\prime,yb\beta^\prime}
\mbox{Tr}[U_{kl}(y) \sigma^b] \sigma_{kl}^{\beta^\prime\beta}\right\rangle,
\end{equation}
is a matrix in Dirac space with two independent components~\cite{FaetAl}:
\begin{equation}
C^{\alpha\beta}_{\tilde{g} g}(x_0-y_0)
= C_{1}(x_0-y_0) \delta^{\alpha\beta} + C_{\gamma_0}(x_0-y_0)
\gamma_0^{\alpha\beta}\,,
\end{equation}
with $ C_{1}=\Tr_D[C_{\tilde{g} g}]/4$ and $C_2=\Tr_D[\gamma_0 C_{\tilde{g}
g}(x)]/4$. We see agreement in the masses extracted from each component, see
Fig.~\ref{fig:meff-gluino-glue}. For the final estimates we choose the time
antisymmetric component $C_{1}$, which appears to provide better plateaus.
We apply APE smearing for the links and Jacobi smearing for the fermion
fields in order to optimise the signal-to-noise ratio and to obtain an
earlier plateau in the effective mass.

\subsection{Massless gluino limit}
\label{subsec:massless}

Of high interest in lattice simulations of SYM is the point corresponding to
a massless gluino, where supersymmetry is expected to emerge in the
continuum limit. With Wilson fermions this point must be located by a tuning
procedure due to the additive renormalisation of the bare gluino mass.

The subtracted gluino mass can be determined in different ways. It can be
directly obtained from the study of lattice SUSY Ward-Identities (WIs) as
discussed in~\cite{FaetAl}. We have implemented the determination of the
necessary operators for the WIs. Apart from confirming the smallness of
lattice corrections to the WIs, consistent with $\mathcal{O}(a)$ effects,
they give results for the gluino mass up to a renormalisation factor. On the
other hand, the point of vanishing gluino mass can be estimated in an
indirect way from the vanishing of the adjoint pion mass. Indeed, as
mentioned above, the pion mass squared $(am_{\pi})^2$ is expected to vanish
linearly with the (renormalised) gluino mass.

\begin{figure}[!htb]
\vspace{1.6cm}
\centering
\includegraphics[width=0.7\textwidth]{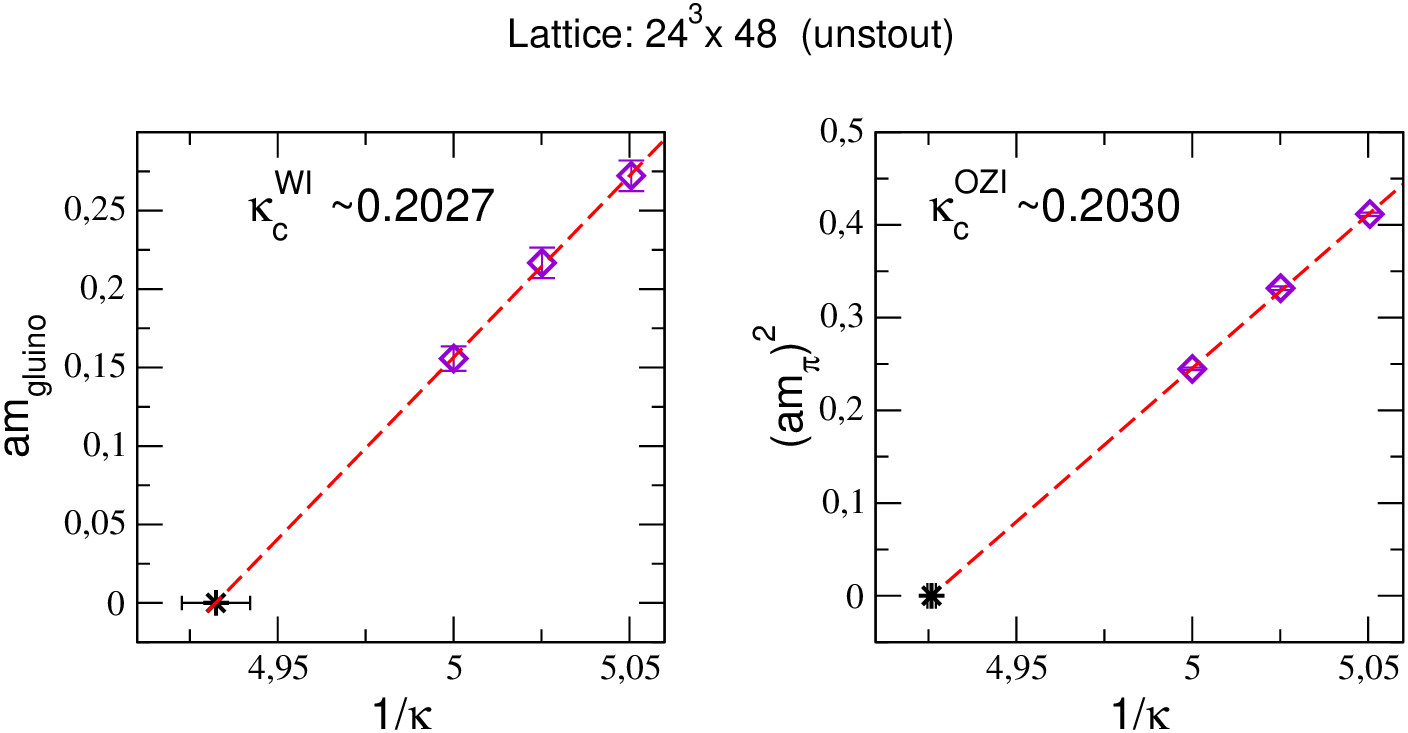}
\parbox[t]{0.8\textwidth}{%
\caption{\label{fig:wi}
The gluino mass from the SUSY Ward identities (left panel) and the pion mass
squared (right panel) as function of the inverse hopping parameter
$1/\kappa$.  The critical value $\kappa_c$ is indicated by the asterisk
symbol.}
}
\end{figure}

Both the WIs and adjoint pion mass methods give consistent estimates of the
critical hopping parameter $\kappa_{c}$ corresponding to vanishing gluino
mass.  As an example, in Fig.~\ref{fig:wi} we show the the gluino mass and
the pion mass squared as a function of $1/\kappa$ for the runs with unstout
links on the $24^3 \cdot 48$ lattice. Both clearly show a linear behaviour.
The linear extrapolations to vanishing gluino mass give $\kappa_{c}^{WI} =
0.2027(4)$ from the Ward identities and $\kappa_{c}^{OZI} = 0.20300(5)$ from
the pion mass. Similarly, for the runs with stout links on the $24^3 \cdot
48$ lattice we obtain $\kappa_{c}^{WI} = 0.15883(85)$ from the Ward
identities and $\kappa_{c}^{OZI} = 0.15793(4)$ from the pion mass, which
agree within errors.

%

\section{Spectrum of bound states}
\label{sec:spectrum}

\begin{figure}[!htb]
\vspace{.8cm}
\centering
\includegraphics[width=0.9\textwidth]{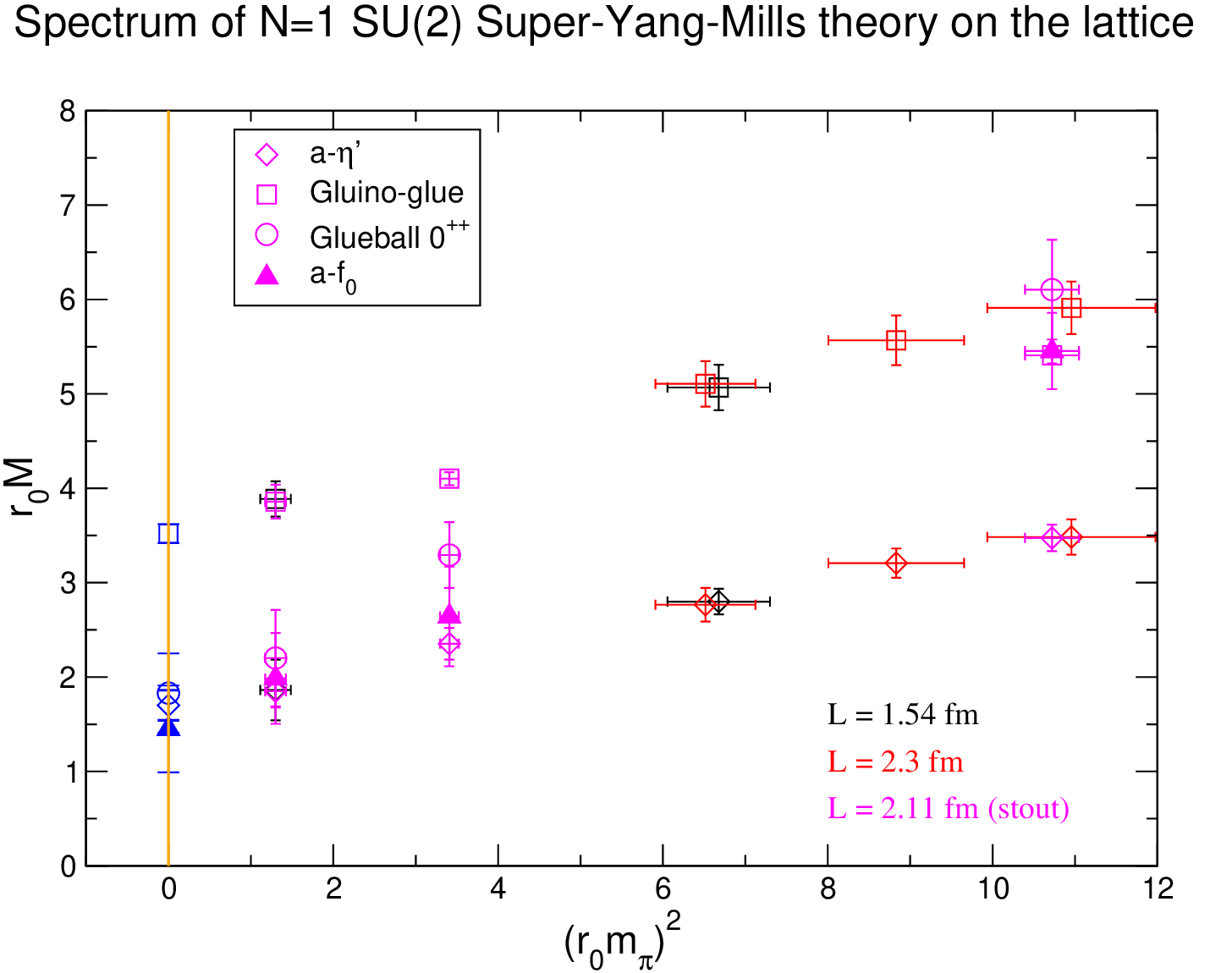}
\parbox[t]{0.8\textwidth}{%
\caption{\label{fig:spectrum}
Low-lying bound state masses of ${\cal N}=1$ SU(2) SYM as a function of the 
adjoint pion mass squared in physical units. The blue symbols represent the
extrapolations to the massless gluino limit.}
}
\end{figure}

The masses of the lightest bound states of low-energy $\mathcal{N}=1$ SYM
determined in this work are collected in Table~\ref{tab:spectrum} and a
graphic representation is shown in Fig.~\ref{fig:spectrum}.
\begin{table}[!ht]
\caption{\em Results for the low-lying bound state masses of ${\cal N}=1$ SU(2) 
SYM from the various runs. The masses are given in lattice units.}
\label{tab:spectrum}
\begin{center}
\begin{tabular}{lll|llll}
Run & $L$ & $\kappa$ & $a$-$\eta^\prime$ & $a$-$f_{0}$ & $\tilde{g}g$
& glub. $0^{++}$ \\
\hline\hline
$A$ & 16 & 0.1900 & 1.3115(67) & 2.229(80) & 1.862(21) & 1.291(13)\\
$B$ & 16 & 0.1800 & 1.0396(72) & 1.27(18) & 1.546(14) & 1.156(51)\\
$C$ & 16 & 0.2000 & 0.5425(71) & 0.931(53) & 0.982(10) & 0.941(18)\\
$D$ & 16 & 0.2000 & 0.361(60) & 0.87(10) & 0.7532(96) & 0.819(19)\\
\hline
$\rule{0pt}{14pt}\bar A$ & 24 & 0.1980 & 0.675(18) & 1.15(12) & 1.1456(82) & \\
$\bar B$ & 24 & 0.1990 & 0.6215(86) & 1.314(32) & 1.0789(95)  &   \\
$\bar C$ & 24 & 0.2000 & 0.536(24) & 0.863(81) & 0.9895(70) & 0.781(24)\\
\hline
$A_{s}$ & 24 & 0.1500 & 1.0114(82) & 1.07(16) & 1.302(14) & \\
$B_{s}$ & 24 & 0.1550 & 0.614(23) & 0.964(70) & 0.9559(48) & 1.079(92)\\
$C_{s}$ & 24 & 0.1570 & 0.416(29) & 0.467(93) & 0.7250(56) & 0.582(61)\\
$D_{s}$ & 24 & 0.1575 & 0.327(30) & 0.351(85) & 0.682(30) & 0.389(90)\\
\hline
\end{tabular}
\end{center}
\end{table}

The masses in Fig.~\ref{fig:spectrum} are multiplied by the extrapolated
value of the Sommer scale parameter and plotted as a function of the squared
adjoint pion mass for $(r_{0}m_{\pi})^2 < 12$. The lightest simulated
adjoint pion mass is about 440 MeV in our units.  The vertical line in
Fig.~\ref{fig:spectrum} indicates the massless gluino limit where SUSY
restoration is expected up to $\mathcal{O}(a)$ effects.  The physical extent
of the lattice is $1.5 - 2.3$\,fm.

The bound state masses appear to be characterised by a linear dependence on
$(r_{0}m_{\pi})^2$, in accordance with the prediction of
\cite{EvansHsuSchwetz}. An extrapolation of our data with stout links
(points $B_s$ to $D_s$), which have better numerical quality than the
unstout ones, to the massless gluino limit yields the numbers in Table
\ref{tab:extrapol}.
\begin{table}[!ht]
\caption{\em Bound state masses in physical units 
extrapolated to the massless gluino limit.}
\label{tab:extrapol}
\begin{center}
\begin{tabular}{c|cccc}
& $a$-$\eta^\prime$ & $a$-$f_{0}$ & $\tilde{g}g$ & glub. $0^{++}$ \\
\hline
$m$ [MeV] & 670(63) & 571(181) & 1386(39) & 721(165)
\end{tabular}
\vspace*{-0.7em}
\end{center}
\end{table}

The gluino-glueball $({\tilde g}g)$ with a mass of about 1386 MeV turns out
to be considerably heavier than the $a$-$\eta^\prime$ with a mass of 670
MeV. Furthermore, the masses of the scalar glueball and the scalar meson
$a$-$f_{0}$ are near the mass of the pseudoscalar $a$-$\eta^\prime$. The
behaviour of scalars is compatible with mixing between $0^+$ glueball and
$a$-$f_{0}$. The pattern of scalar masses suggests a lower supermultiplet,
while the spin-1/2 candidate remains heavier up to the smallest simulated
gluino mass in this simulation, and also after extrapolation to $\kappa_c$.
Whether this outcome is a discretisation artefact or a physical effect, as
claimed in \cite{BeMi}, should become clear in future studies at finer
lattice spacings. As the data at small gluino mass are preliminary, it would
be premature to make judgements about this point.

\section{Conclusion}
\label{sec:conclusion}

In this work first {\em quantitative} results on the low-energy spectrum of
${\cal N}=1$ supersymmetric Yang-Mills theory are obtained. Physical volumes
larger than 2 fm have been simulated, which is the volume usually required
for spectrum studies in lattice gauge theory.  The comparison of masses on
different volumes in otherwise same conditions reveals negligible finite
size effects at least for moderate gluino masses. We have collected higher
statistics and have used efficient dynamical algorithms such as TS-PHMC,
which is suitable for light fermion masses. In addition, the supersymmetric
Ward identities and other observables like the confinement potential have
been investigated.

From the results of the mass spectrum the question of the gluino-gluino and
gluino-glueball mass splitting remains open. It can only be answered by
further simulations allowing an extrapolation to the continuum limit.
\vspace*{1.5em}

\noindent
{\large\bf Acknowledgements}

\noindent
This work has been supported by the German Science Foundation (DFG) under
contracts Mu757/9 and Mu757/13, and by the John von Neumann Institute for
Computing (NIC) with grants of computing time.  K.D.\ would like to thank
the German Academic Exchange Service (DAAD) for support. The numerical
simulations of this work have been performed on the Blue Gene L/P and JuMP
systems at JSC J\"ulich, Opteron PC-cluster at RWTH Aachen and the ZIV
PC-cluster of the university of M\"unster.


\end{document}